\pdfoutput=1
\documentclass[notitlepage,nofootinbib,aps]{revtex4-1}
\usepackage{amssymb,amsmath,graphicx,color,microtype}
\usepackage[section]{placeins}

\begin{document}
\title{Inflation and Alternatives with Blue Tensor Spectra}
\author{Yi Wang}
\email{yw366@cam.ac.uk}
\affiliation{Centre for Theoretical Cosmology, DAMTP, University of Cambridge, Cambridge CB3 0WA, UK}

\author{Wei Xue}
\email{wei.xue@sissa.it}
\affiliation{INFN, Sezione di Trieste, SISSA, via Bonomea 265, 34136 Trieste, Italy}

\begin{abstract}
  We study the tilt of the primordial gravitational waves spectrum. 
  A hint of blue tilt is shown from analyzing the BICEP2 and POLARBEAR data.
  Motivated by this, we explore the possibilities of blue tensor spectra from 
   the very early universe cosmology models, 
   including null energy condition violating 
   inflation, inflation with general initial conditions, and 
   string gas cosmology, etc. For the simplest G-inflation, 
   blue tensor spectrum also implies blue scalar spectrum. 
   In general, the inflation models with blue tensor spectra indicate 
   large non-Gaussianities. On the other hand, string gas cosmology predicts blue tensor spectrum with highly Gaussian fluctuations. 
   If further experiments do confirm the blue tensor 
   spectrum, 
   non-Gaussianity becomes a distinguishing test between 
   inflation and alternatives.
\end{abstract}

\maketitle

\section{Introduction}

Inflation \cite{Guth81, Linde82} is the leading paradigm for the very early universe cosmology. Inflation has been proposed to explain the horizon, flatness and monopole problems in the standard hot big bang cosmology, and almost all the predictions of the simplest inflation model have now been tested. 
The observational tests of inflation includes

\begin{itemize} 
\item Coherent and nearly scale invariant power spectrum of density perturbations. The power spectrum of the simplest slow roll inflation is \cite{Planck16}
\begin{align}
  P_\zeta = \frac{H^2}{8\pi^2\epsilon M_p^2} \simeq 2.43 \times 10^{-9}~. 
\end{align}

\item A small tilt of the scalar power spectrum.
\begin{align}
  n_s - 1 = -2\epsilon - \eta \simeq 0.96~,
\end{align}
where $\eta \equiv \dot \epsilon / (H\epsilon)$ is the slow roll parameter defined from expansion.
Now $n_s\geq 1$ is ruled out under the assumptions of simplest inflation models.

\item Nearly Gaussian density fluctuations. The non-Gaussianities of the density fluctuations are tightly constrained at
\begin{align}
  f_\mathrm{NL}^\mathrm{local} = 2.7 \pm 5.8~, \qquad f_\mathrm{NL}^\mathrm{equil} = -42\pm 75
\end{align}
for the local shape and equilateral shape non-Gaussianities respectively. Those numbers indicate that, non-Gaussian components of the primordial fluctuations, even if exist, have to be at least 3$\sim$4 orders of magnitudes smaller than the Gaussian component.

\item Gravitational waves. The recent BICEP2 experiment reports an over $5\sigma$ detection of gravitational waves \cite{Ade:2014xna}, with tensor to scalar ratio \footnote{There is a debate about whether the observed polarisation signal comes from primordial gravitational waves or dust contamination \cite{Mortonson:2014bja, Flauger:2014qra}.}
\begin{align}
  r = 0.20^{+0.07}_{-0.05} ~(1\sigma \mathrm{CL})~.
\end{align}
This corresponds to a gravitational wave fluctuation amplitude
\begin{align}
  P_\mathrm{T} = \frac{2H^2}{\pi^2 M_p^2} = 4.8 \times 10^{-10}~.
\end{align}
\end{itemize}

Despite of the great success of inflation, there are still a few outstanding challenges for theorists and experimentalists. 

On the theoretical side, large field inflation is now favored. However, large field inflation is hard to construct from the effective field theory, and stringy UV completion points of view. The UV completion of inflation has long suffered from an $\eta$-problem \cite{Copeland:1994vg}, in which the mass of the inflaton is theoretically too large to allow enough e-folds of inflation. However, with the current data, a more serious $\epsilon$-problem emerges -- the observed energy scale of inflation is too high for an effective field theory or stringy model building to be under control. For single field slow roll inflation, at every e-fold, the inflaton rolls a distance of order $0.1 M_p$. In perturbative string theory, this field motion per e-fold is comparable with, or greater than the string scale $M_s$. As a result, one can no longer safely globally expand the inflaton field and ignore non-renormalizable terms. More discussions and a local reconstruction of the inflationary potential can be found in \cite{Ma:2014vua}

On the observational side, there is yet another (and maybe the last in the foreseeable future, unless nature is so kind as to imprint other relics on the CMB sky or in the large scale structure) test for inflation which is possible in light of BICEP2, but not yet achieved -- the tilt of the tensor power spectrum. The simplest inflation models predict a consistency relation between $n_\mathrm{T}$ and $r$ as
\begin{align}
  n_\mathrm{T} = - \frac{r}{8} = -0.025~.
\end{align}
Currently the data has not been good enough to test $n_\mathrm{T}$ precisely. However, there are a lot of ongoing and upcoming experiments in the near future \cite{Planck22, Ade:2014afa, Austermann12, Niemack10, Eimer12}, measuring $r$ at different scales, which provides a possibility for a precise measurement of $n_\mathrm{T}$.

In this paper, we shall explore the possibility of blue $n_\mathrm{T}$. In Section \ref{sec:hints-blue-tensor}, the bound on $n_\mathrm{T}$ is derived the BICEP2 \cite{Ade:2014xna} and POLARBEAR \cite{Ade:2014afa} data. The string gas cosmology, null energy condition (NEC) violating inflation, and general initial condition are addressed in Sections \ref{sec:string-gas-cosmology}, \ref{sec:infl-viol-nec} and \ref{sec:infl-gener-init} respectively. In Section \ref{sec:other-possibilities}, a few other possibilities are discussed, including external sources for tensor modes, modified gravity and matter bounce. We conclude in Section \ref{sec:concl-disc}.

\section{Hints for blue tensor spectra}
\label{sec:hints-blue-tensor}

In this section, we show the constraints on $r$ and $n_\mathrm{T}$ from BICEP2, \textit{Planck} and POLARBEAR data. The public codes CosmoMC \cite{Lewis:2002ah} and CLASS \cite{Blas:2011rf} are used in the calculation. Here we vary $r$ and $n_\mathrm{T}$ in the Boltzmann codes, and other cosmological parameters $\{\Omega_bh^2, \Omega_ch^2, \theta, \tau, n_s, \log A\}$ are set at the best fit value from Planck results \cite{Hinshaw:2012aka} where not explicitly mentioned.

\begin{figure}[!htb!]
   \centering
   \hspace{-2.7 cm} 
   \begin{minipage}{0.3\textwidth}
   \includegraphics[scale=0.45]{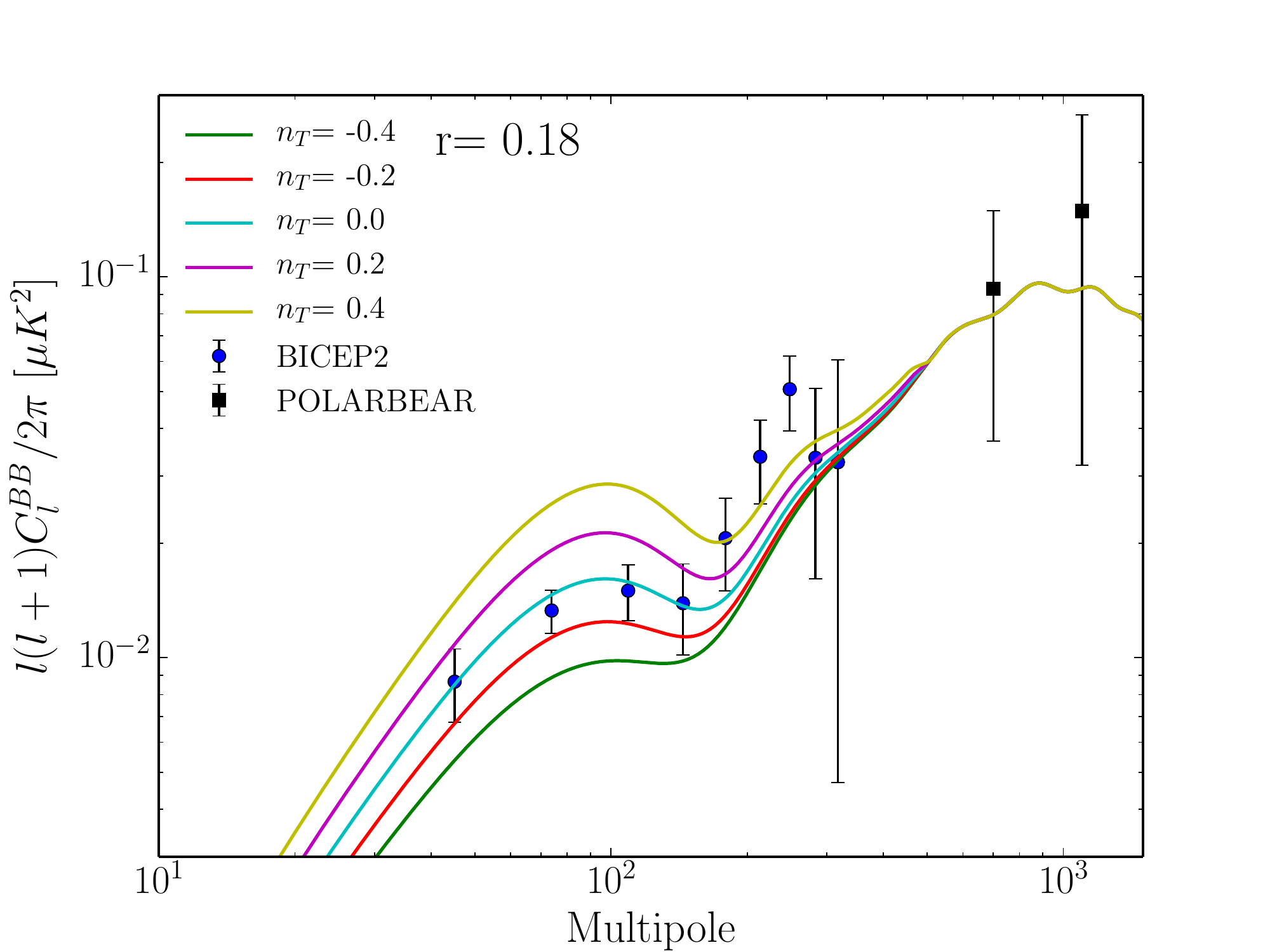}
    \end{minipage}%
   \hspace{3.0 cm} 
  \begin{minipage}{0.3\textwidth}
   \includegraphics[scale=0.45]{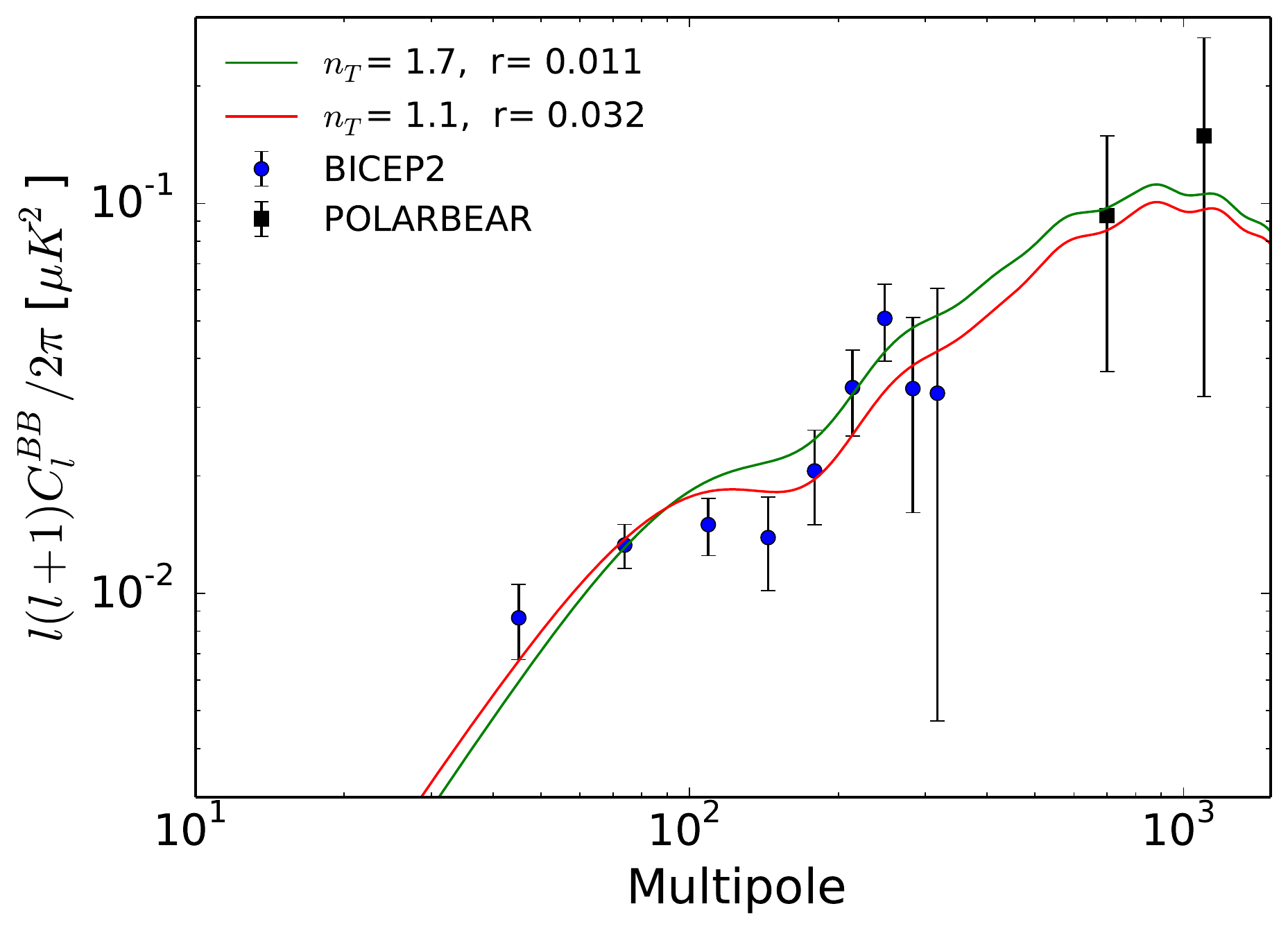}
    \end{minipage}
  \caption{ The simulated $BB$ power spectrum with different $r_{0.002}$ and $n_\mathrm{T}$. Left
      panel: $r_{0.002}$ is fixed and variation of $n_\mathrm{T}$ is illustrated. Right panel: some
         sets of $r_{0.002}$ and $n_\mathrm{T}$ are chosen to get good fit against data. 
         In both panels $r$ is calculated at pivot scale $k=0.002$ Mpc$^{-1}$.}
\label{fig:rnt_sample}
\end{figure}

The relation between $r$ and $n_\mathrm{T}$ is illustrated in fig.~\ref{fig:rnt_sample}.
At $r = 0.18$, the flat tensor spectrum gives a good fit to the data, which 
is consistent with the result of the BICEP2 \cite{Ade:2014xna}. With the fixed
ratio $r$ at pivot scale $0.002 ~ \mathrm{Mpc}^{-1}$, by increasing or decreasing
the tilt $n_\mathrm{T}$ by $0.2$, it starts to be ruled out by the current data.
If we consider smaller tensor-to-scalar ratio,
indicated from the right panel of fig.~\ref{fig:rnt_sample},
a blue tensor spectrum\footnote{As we shall see in eqs. \eqref{eq:b2best} and \eqref{eq:b2p13best}, the values illustrated in the figure corresponds to best fit values of BICEP2 and BICEP2 + Planck respectively. } is required by the BICEP2 data. One can observe from the right panel of fig.~\ref{fig:rnt_sample} that very blue $n_\mathrm{T}$ fits all data points within $2\sigma$. This is different from the $n_\mathrm{T}=0$ case, where some data points are too high and may be considered as outliers. 

\begin{figure}[!htb!]
   \centering
   \hspace{-2.7 cm} 
   \begin{minipage}{0.3\textwidth}
   \includegraphics[scale=0.45]{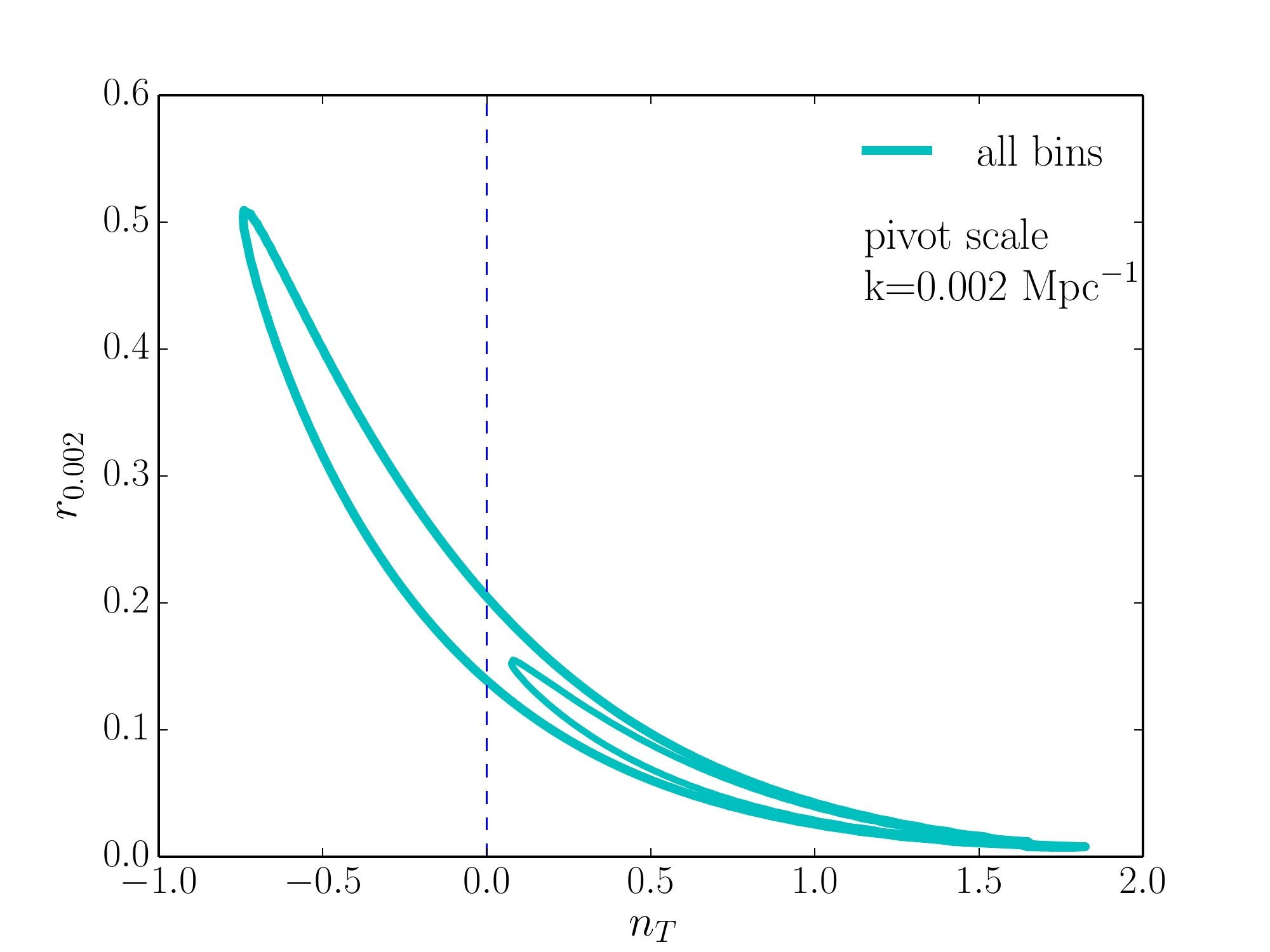}
    \end{minipage}%
   \hspace{3.0 cm} 
  \begin{minipage}{0.3\textwidth}
   \includegraphics[scale=0.45]{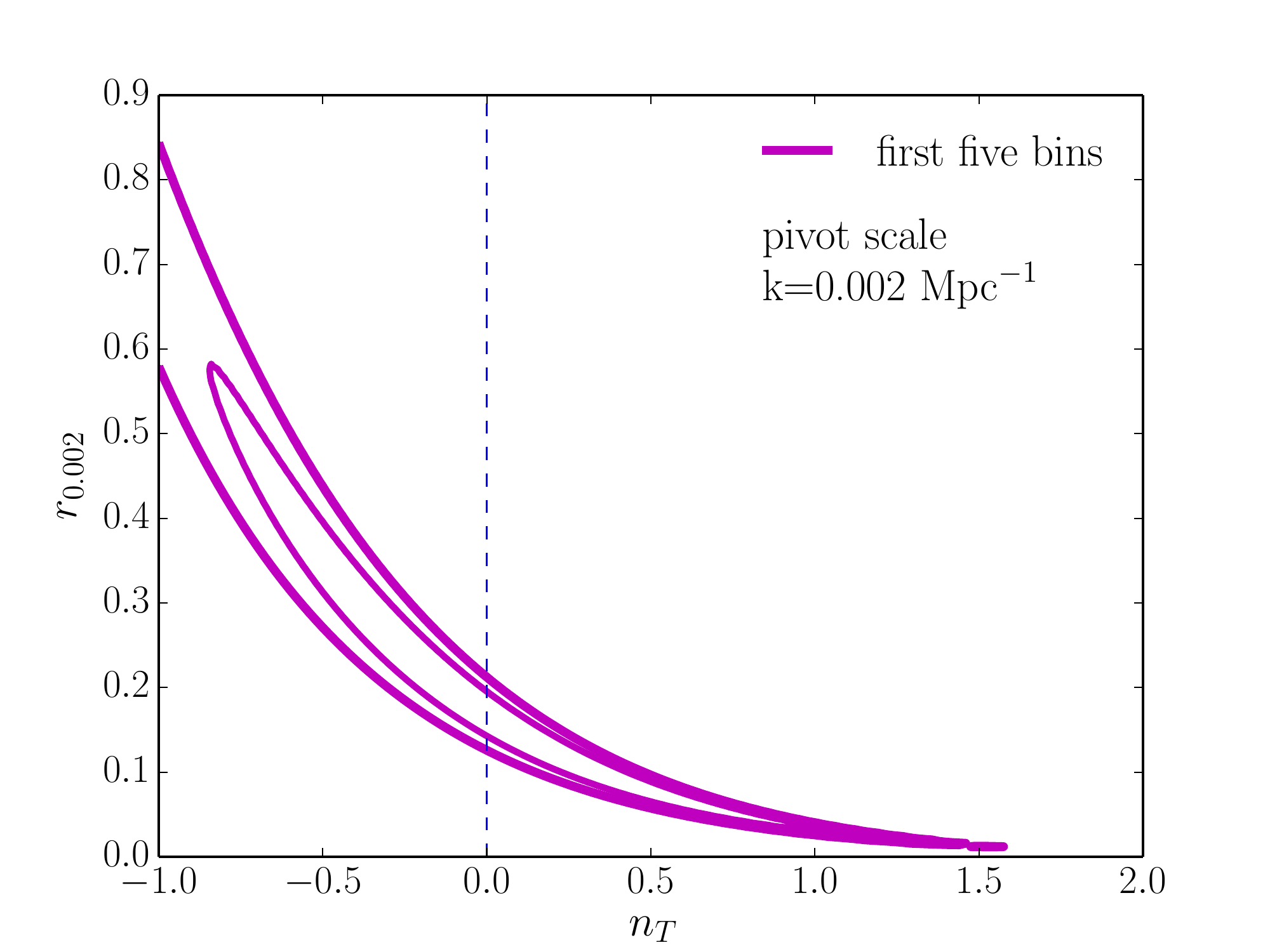}
    \end{minipage}
  \caption{ Contour plot of $r_{0.002}$ vs $n_\mathrm{T}$ to fit against the data.
   Left panel: fit the data of the BICEP2 and POLARBEAR.
   Right panel: fit the first five data bins of BICEP2 and 
      treat the others as upper bound. Note that the strong correlation between $n_t$ and $r_{0.002}$ is an artifact from the choice of pivot scale. We have chosen $k=0.002 \mathrm{Mpc}^{-1}$ to match with BICEP2 conventions. However, the BICEP2 experiments measures $k\sim 0.01 \mathrm{Mpc}^{-1}$. Thus given a similar tensor spectrum at $k\sim 0.01 \mathrm{Mpc}^{-1}$, the large tilt modifies the $r_{0.002}$ value significantly.}
\label{fig:rnt_contour}
\end{figure}

A more precise correlation of $n_\mathrm{T}$ and $r$ is illustrated in fig.~\ref{fig:rnt_contour}, where the pivot scale at $0.002~\mathrm{Mpc}^{-1}$ is the same as the analysis of BICEP2 paper~\cite{Ade:2014xna}.

However, as one can observe from fig.~\ref{fig:rnt_contour}, $r$ and $n_\mathrm{T}$ are strongly correlated when the pivot scale is chosen at $0.002~\mathrm{Mpc}^{-1}$. This makes detailed analysis inconvenient. The strong correlation comes from the fact that the BICEP2 experiment is actually not observing the B-mode signals at their pivot scale $0.002~\mathrm{Mpc}^{-1}$, but instead the observation is made at about $0.01~\mathrm{Mpc}^{-1}$. This does not matter for the BICEP2 data analysis itself because $n_\mathrm{T}=0$ is chosen in their data analysis. However, in our case, when $n_\mathrm{T}$ is allowed as a free parameter, we had better pay more attention to the pivot scale. In the remainder of this section, we shall choose the pivot scale at $0.01~\mathrm{Mpc}^{-1}$ and calculate $r$ at this pivot scale.

In table~\ref{tab:chi2}, the best fit parameters and their likelihood are summarized. One can observe that from the BICEP2 data only, including $n_\mathrm{T}$ improves $\chi^2$ by 3.161. Considering that one additional parameter is induced, it is useful compute the Akaike information criterion (AIC) or the Bayesian information criterion (BIC) \footnote{Alternatively, the Bayesian information criterion may be applied. But we choose to apply AIC because BIC requires the number of observed data points related to the fitting. However, only part of the data points we have used here are sensitive to the additional parameter $n_\mathrm{T}$. The BICEP2 E-mode polarization, the WMAP polarization and the high $\ell$ part of the Planck temperature data has a large amount of data points which are not sensitive to the change of $n_\mathrm{T}$. Thus we choose AIC to compare between models.} to see how much improvements $n_\mathrm{T}$ would bring. AIC can be calculated as
\begin{align}
  \mathrm{AIC} = 2k + \chi^2 ~, 
\end{align}
where $k$ is the number of model parameters. Thus including $n_\mathrm{T}$ results in $\Delta\mathrm{AIC} = 1.161$. The number not very significant, though showing hint that blue $n_\mathrm{T}$ would fit data slightly better. 
To explore the goodness-of-the-fit, we can also look at the relative $\chi^2$ 
and p-value of
the BICEP2 BB correlation. Including BICEP2 BB correlation and POLARBEAR,
the relative  $\chi^2_{rel} =1.01 $, p-value is $0.429$ in the case of $n_\mathrm{T}$ 
as a free parameter and 
$\chi^2_{rel} = 1.14$, p-value is $0.317$ for $n_\mathrm{T}=0$. Both of $\chi^2_{rel}$ are 
close to $1$ meaning both of them are good fit, but the one with the parameter $n_\mathrm{T}$ is slightly better.
The improvement mainly comes from the small scale excess of the B-mode power spectrum \footnote{Note that the small scale excess is not observed in the BICEP2 $\times$ Keck (preliminary) cross correlation \cite{Ade:2014xna}. However, in the BICEP2 $\times$ Keck (preliminary) cross correlation, the large scale data points are lower than the theoretical prediction, again showing some evidence of blue $n_\mathrm{T}$.}.

However, the significance of blue $n_\mathrm{T}$ increase dramatically once Planck data is considered. After allowing to vary $n_\mathrm{T}$, $\chi^2$ is improved by $11.608$ for this single parameter. Namely, the fitting of $\ell \leq 49$ part of Planck temperature data is improved by $\Delta\chi^2 = 6.637$. The fitting of $\ell \geq 50$ part of Planck temperature data is improved by $\Delta\chi^2 = 0.703$. The fitting of BICEP2 data is improved by 4.945. On the other hand, blue $n_\mathrm{T}$ actually fit slightly worse for WMAP polarization, by $\Delta\chi^2 = 0.677$. As a result, one can calculate that $\Delta \mathrm{AIC} = 9.608$, showing strong evidence of blue $n_\mathrm{T}$.

It is straightforward to understand why inclusion of Planck data strongly prefers blue $n_\mathrm{T}$. Note that among the likelihoods listed in table~\ref{tab:chi2}, the most significant improvement comes from the $\ell \leq 49$ part of Planck temperature data. This is because, there is already an anomalous temperature power deficit in the small $\ell$ part of the Planck data, by an amount of $5\% \sim 10\%$, at the statistical significance of $2.5\sim 3\sigma$. On the other hand, the existence of tensor mode, when assuming $n_\mathrm{T}=0$, adds to the theoretically predicted temperature power by another $5\% \sim 10\%$, which fits data even worse.

By inclusion of $n_\mathrm{T}$, the above tension between Planck and BICEP is reconciled \footnote{This point is added since v2 of this paper. A similar viewpoint \cite{Ashoorioon:2014nta} has already appeared before our revision.}. As one can compare to the last column of table~\ref{tab:chi2}, after adding blue $n_\mathrm{T}$, the likelihood of Planck low $\ell$, Planck high $\ell$ and WMAP polarization are all comparable with the best fit values before BICEP2. Thus the tension between Planck and BICEP2 can be reconciled by blue $n_\mathrm{T}$.

\begin{figure}[htbp]
  \centering
  \includegraphics[width=\textwidth]{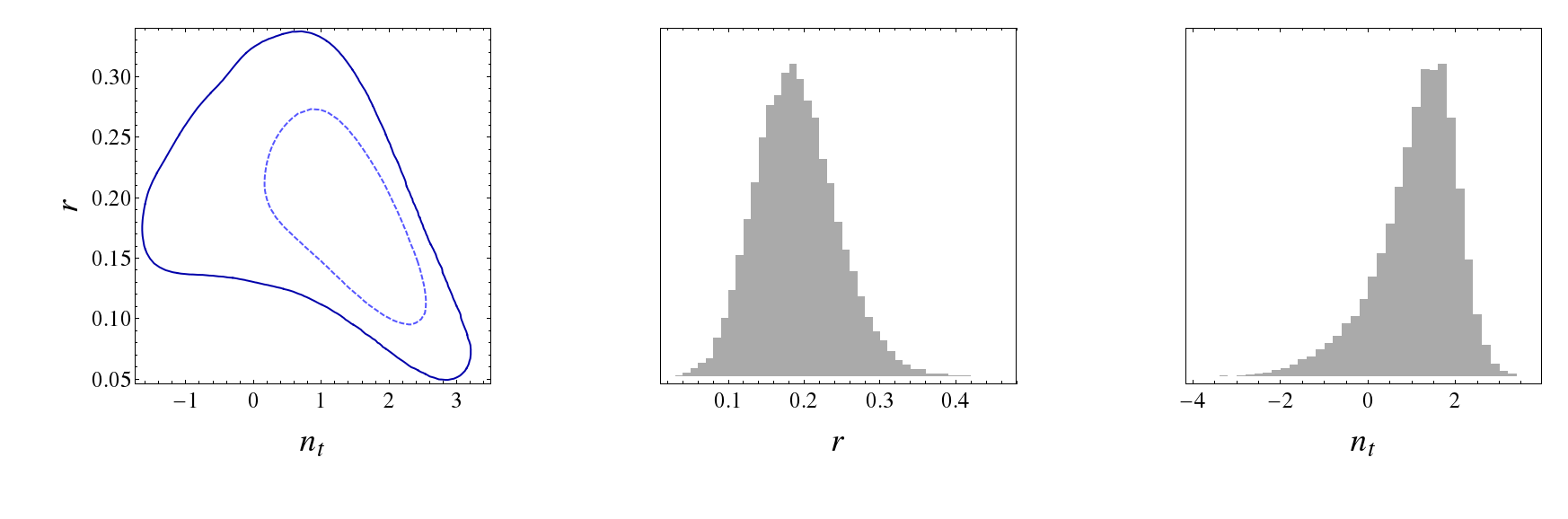}
  \caption{\label{fig:b9plot} The $r-n_\mathrm{T}$ contour (left panel), and the likelihood for $r$ (middle panel) and $n_\mathrm{T}$ (right panel) from BICEP2 (9 bins).}
\end{figure}

\begin{figure}[htbp]
  \centering
  \includegraphics[width=\textwidth]{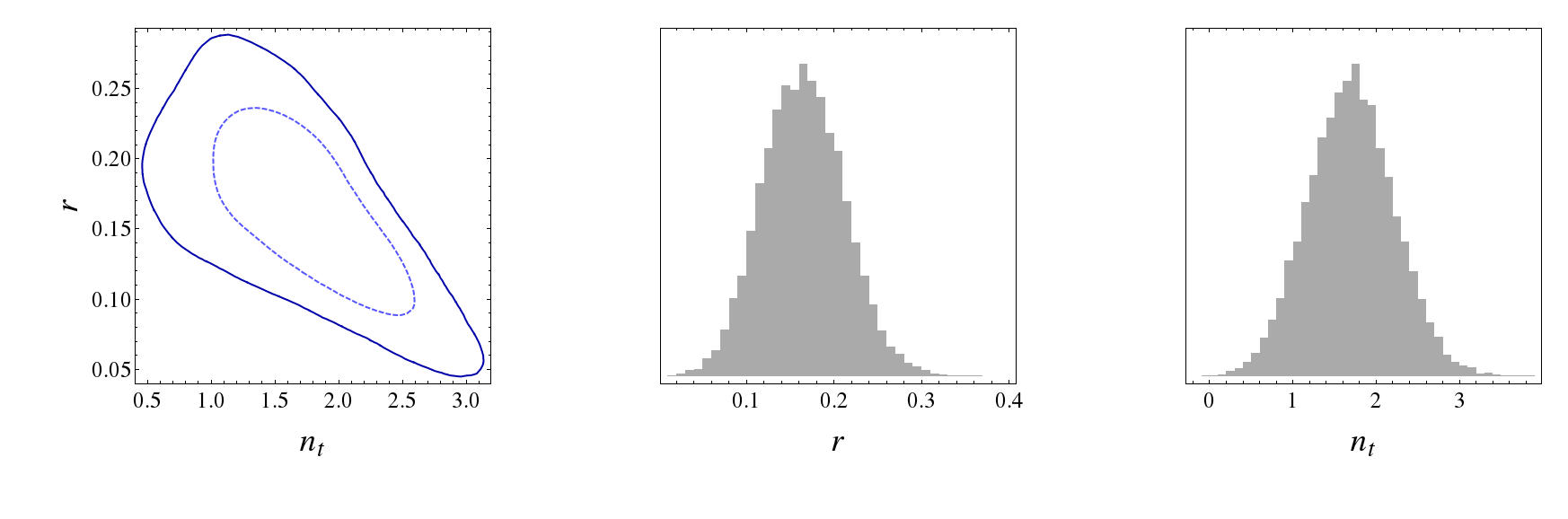}
  \caption{\label{fig:b9pwplot} The $r-n_\mathrm{T}$ contour (left panel), and the likelihood for $r$ (middle panel) and $n_\mathrm{T}$ (right panel) from BICEP2 (9 bins) + Planck (2013) + WMAP polarization.}
\end{figure}

To further get constraints of $r$ and $n_\mathrm{T}$, we run MCMC chains to get samples according to the likelihood code of BICEP2, Planck and WMAP polarization. The $1\sigma$ and $2\sigma$ contours for BICEP2 only (9 bins) is plotted in fig.~\ref{fig:b9plot}, and BICEP2+Planck+WP is plotted in fig.~\ref{fig:b9pwplot}. Again one observes that blue $n_\mathrm{T}$ is preferred. The mean value and standard derivation of $r$ and $n_\mathrm{T}$ from BICEP2 (9 bins) is
\begin{align}\label{eq:b2best}
  r = 0.19 \pm 0.05 ~, \qquad n_\mathrm{T} = 1.10 \pm 0.93~.
\end{align}
The constraints from BICEP2 (9 bins) + Planck (2013) + WMAP polarization is
\begin{align}\label{eq:b2p13best}
  r = 0.17 \pm 0.05 ~, \qquad n_\mathrm{T} = 1.70 \pm 0.52~.
\end{align}

\begin{figure}[htbp]
  \centering
  \includegraphics[width=\textwidth]{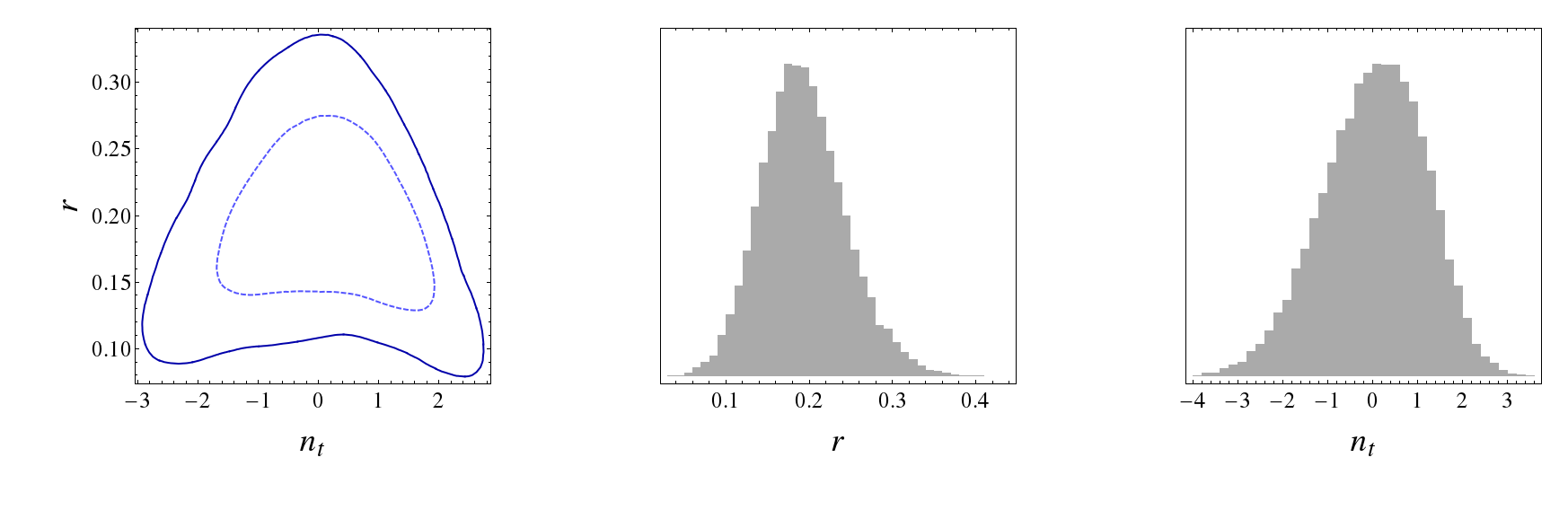}
  \caption{\label{fig:b5plot} The $r-n_\mathrm{T}$ contour (left panel), and the likelihood for $r$ (middle panel) and $n_\mathrm{T}$ (right panel) from BICEP2 (5 bins).}
\end{figure}

On the other hand, it is worth noticing that if one only take the first 5 bins from BICEP2 data, there is no evidence of blue $n_\mathrm{T}$. This is plotted in fig.~\ref{fig:b5plot} for comparison purpose. Because as we have discussed, the two major sources of blue $n_\mathrm{T}$ are the observed small scale excess of the B-mode power spectrum of BICEP2, and the observed large scale suppression of the temperature spectrum of Planck. None of those are present in the first 5 bins of BICEP2 data. The constraints from those 5 bins are
\begin{align}
  r = 0.19 \pm 0.05 ~, \qquad n_\mathrm{T} = 0.030 \pm 1.14~.
\end{align}

\begin{table}[htbp]
  \begin{center}
    \begin{tabular}{ | c || c | c || c | c | c |}
      \hline
                      & B2 ($r$) & B2 ($r$,$n_\mathrm{T}$) & B2+P13+WP ($r$) &  B2+P13+WP ($r$,$n_\mathrm{T}$)
      & P13+WP (without B2)\\ \hline\hline
      $r$             & 0.214   & 0.169         & 0.155          & 0.178      & 0      \\ \hline
      $n_\mathrm{T}$           & 0       & 1.704         & 0              & 1.539      & 0      \\ \hline\hline
      P13 low $\ell$  &      &            & -0.061         & -6.698     & -6.760 \\ \hline
      P13 high $\ell$ &      &            & 7796.327       & 7795.624   & 7795.276 \\ \hline
      WP              &      &            & 2013.469       & 2014.146   & 2014.305\\ \hline
      BICEP2          & 38.599  & 35.438        & 39.899         & 34.954     &      \\ \hline
      Total           & 38.599  & 35.438        & 9849.634       & 9838.026   & 9802.821 (without B2)\\ \hline\hline
      $\Delta\chi^2$           &      & 3.161         &             & 11.608     &      \\ \hline
      $\Delta$AIC          &      & 1.161         &             & 9.608      &      \\ \hline
    \end{tabular}
  \end{center}
  \caption{\label{tab:chi2} The best fit values for $r$ and $n_\mathrm{T}$, likelihood and AIC. The 5 columns corresponds to: [Column 1] BICEP2 only (9 bins) with varying $r$ only, [Column 2] BICEP2 only (9 bins) with varying $r$ and $n_\mathrm{T}$, [Column 3] BICEP2 (9 bins) + Planck (2013) + WMAP polarization with varying $r$ only, [Column 4] BICEP2 (9 bins) + Planck (2013) + WMAP polarization with varying $r$ and $n_\mathrm{T}$, and [Column 5] Planck (2013) + WMAP polarization with $r=0$ (for comparison purpose), respectively. Note that in the first two columns (BICEP2 data only) the other cosmological parameters $\{\Omega_bh^2, \Omega_ch^2, \theta, \tau, n_s, \log A\}$ are fixed because BICEP2 alone does not provide reasonable constraints on those parameters. In the minimization of the last two columns (combining with Planck and WP) those additional cosmological parameters and nuisance parameters are also varying.}
\end{table}

\begin{figure}[htbp]
  \centering
  \includegraphics[width=0.6\textwidth]{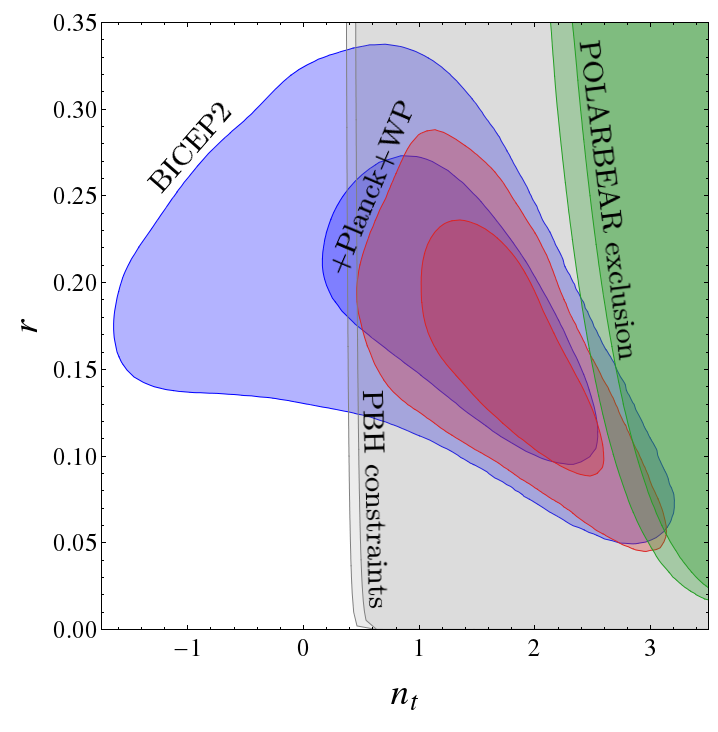}
  \caption{\label{fig:combined} Combined constraints on $r$ and $n_\mathrm{T}$. The blue contours are the $1\sigma$ and $2\sigma$ constraints from BICEP2 (9 bins). The red contours are the constraints from BICEP2 + Planck (2013) + WMAP polarization. The blue shaded region are the $1\sigma$ and $2\sigma$ exclusion curves from POLARBEAR. The gray shaded region is the exclusion curves from PBH constraints, assuming the blue $n_\mathrm{T}$ is also present in smaller scales than CMB observations.}
\end{figure}

Note that when $n_\mathrm{T} \gtrsim 2.5$, the primordial tensor contribution to the B-mode polarization becomes more important than lensing signal even at small scales ($\ell \gg  100$). In this case, the POLARBEAR detection of ``lensing B-mode" becomes a constraint on blue $n_\mathrm{T}$. The $1\sigma$ and $2\sigma$ exclusion curves from POLARBEAR is plotted in fig.~\ref{fig:combined}. On the other hand, the cross-correlation-detection of B-mode from SPTpol does not put constraint on $n_\mathrm{T}$, because the primordial B-mode does not cross correlate with the density perturbation in the simplest inflation models.

Another rather theoretical constraint for blue $n_\mathrm{T}$ is that, very blue tensor tilt cannot last long. Assuming the running of $n_\mathrm{T}$ is not significant, then for $n_\mathrm{T}=1$, it takes about 23 e-folds to bring the tensor mode to be non-perturbative ($P_h\sim 1$). Those non-perturbative tensor modes forms primordial black holes (PBH)\footnote{Those tensor modes may not form primordial black holes directly by themselves, considering the tensor perturbation preserves volume. However, when $P_h\sim 1$, the induced scalar perturbation also becomes of order one, which forms primordial black holes.}, which is constrained from current observations. For $n_\mathrm{T}=2$, it takes about 12 e-folds. Thus assuming
\begin{align}
  \alpha_\mathrm{T} \equiv \frac{d\log n_\mathrm{T}}{d \log k} \ll 1~,
\end{align}
we get $n_\mathrm{T}<0.38$ for 60 e-folds of inflation and $n_\mathrm{T}<0.46$ for 50 e-folds of inflation. Those constraints are plotted in fig.~\ref{fig:combined}. Nevertheless, if $n_\mathrm{T}$ is large, $\alpha_\mathrm{T}$ may be large as well. If $\alpha_\mathrm{T}$ is negative \cite{Cai:2014hja}, it relaxes the above constraint.

Finally, although the current bound on $n_\mathrm{T}$ is far from testing the consistency relation of inflation, it is already informative to disfavor scenarios with very blue or red spectrum. For example, a sharp pulse of gravitational waves with rapid decaying tail towards both ends ($n_\mathrm{T}<-1.5$ or $n_\mathrm{T}>2.5$) are disfavored by the current data.

\section{String gas cosmology}
\label{sec:string-gas-cosmology}

Before a survey of inflationary possibilities for the blue tensor spectra, let us mention alternatives to inflation. This is because the first prediction of a slightly blue $n_\mathrm{T}$ comes from the string gas cosmology \cite{Brandenberger:1988aj, Nayeri:2005ck, Brandenberger:2006xi, Brandenberger:2006vv} (see \cite{Brandenberger:2014faa} for a recent discussion in light of BICEP2 data) 
\footnote{The Ekpyrotic scenario \cite{Khoury:2001wf}(see \cite{Boyle:2003km} for a detailed analysis of tensor modes) also predicts a blue tensor spectrum with $n_\mathrm{T} = 2$.
This is consistent with our $n_\mathrm{T}$ bound if all the BICEP2 data points are fitted. However, the amplitude of the tensor mode is exponentially suppressed on cosmological scales, which cannot be $r=0.2$ at $k=0.002 \mathrm{Mpc}^{-2}$. There is a similar issue for the bouncing Galileons \cite{Qiu:2011cy}. Also, there is recently another model with blue tensor tilt derived from the Hagedorn phase \cite{Biswas:2014kva}.}. On the contrary, for the other models that we shall show below, though they have the possibility to tune the tensor spectrum to be blue, the blueness is not a firm prediction.

In string gas cosmology, the universe was in a string Hagedorn phase before expansion starts. It is conjectured that due to T-duality, and the huge specific heat in the string Hagedorn phase, the universe should stay at a nearly constant temperature (Hagedorn temperature $T_\mathrm{H}$) for a long time, until the string winding modes decays and allow the expansion of the universe (which is also a possible explanation of three large spatial dimensions). 

\begin{figure}[htbp]
  \centering
  \includegraphics[width=0.8\textwidth]{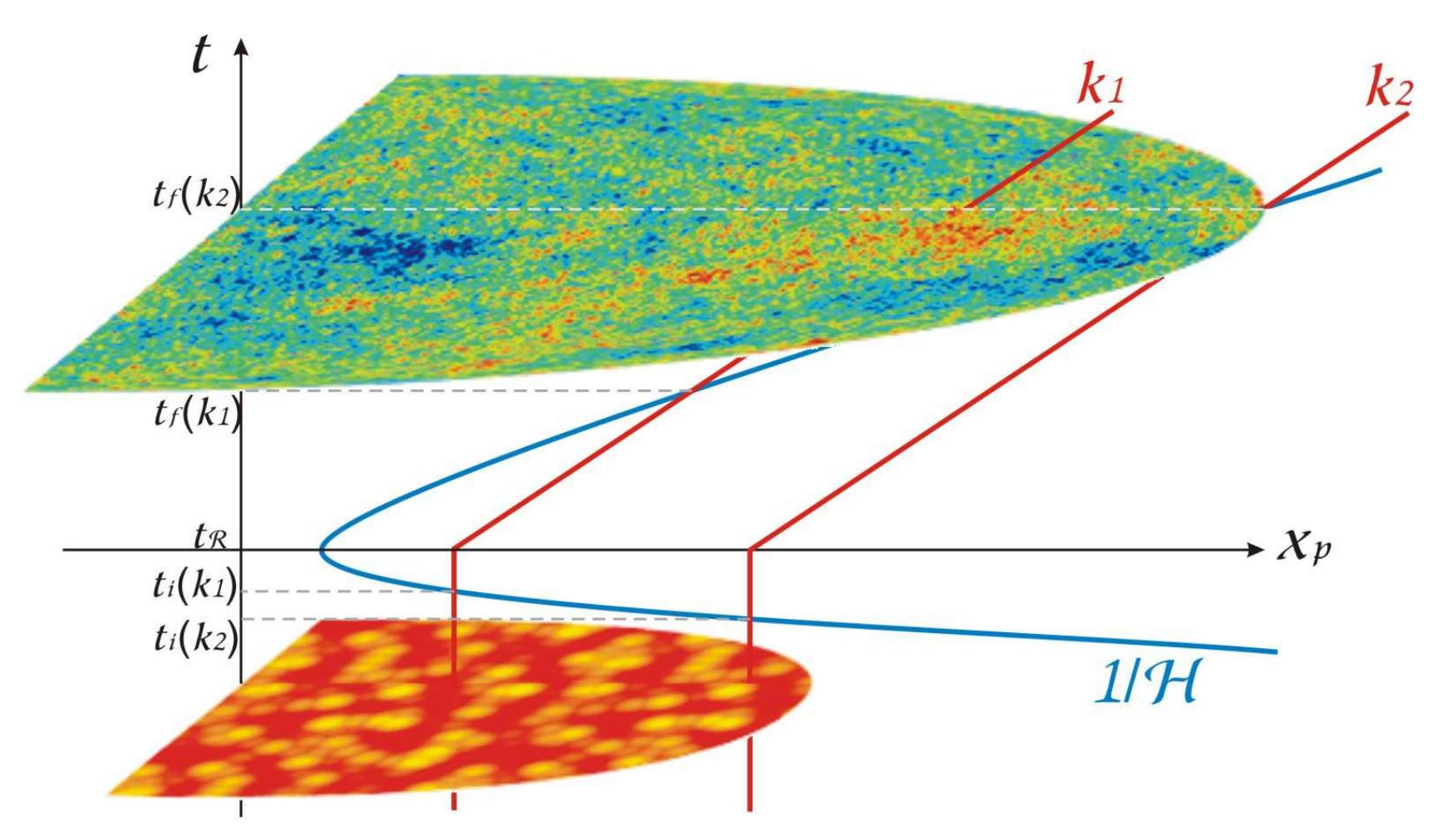}
  \caption{\label{fig:sgc} String gas cosmology. The universe starts from a string Hagedorn phase and a phase transition brings the universe into radiation dominated.}
\end{figure}

The density and tensor fluctuations in string gas cosmology are thermal fluctuations. Those fluctuations were in causal contact in the string Hagedorn phase, and get frozen on super-Hubble scales because of the rapid shrinking of the Hubble horizon (Fig.~\ref{fig:sgc}). The scalar and tensor fluctuations can be calculated as \cite{Nayeri:2005ck, Brandenberger:2006xi, Brandenberger:2006vv}
\begin{align}
  P_\Phi(k) = \left( \frac{l_\mathrm{p}}{l_\mathrm{s}} \right)^4 \frac{T(k)}{T_\mathrm{H}} \frac{1}{1-T(k)/T_\mathrm{H}}~,  
\end{align}
where $\Phi$ is the Newtonian potential. The scalar spectral index is
\begin{align}
  n_s - 1 = - \frac{d \ln (1-T(k)/T_\mathrm{H})}{d \ln k} ~.
\end{align}
From the string gas picture, one expects that $T(k)$ is an decreasing function of $k$ and thus string gas cosmology has red scalar spectrum.

The tensor power spectrum is
\begin{align}
  P_h(k) = \left( \frac{l_\mathrm{p}}{l_\mathrm{s}} \right)^4\frac{T(k)}{T_\mathrm{H}}
  \left( 1-\frac{T(k)}{T_\mathrm{H}}\right) \ln^2 \left[ \frac{1-T(k)/T_\mathrm{H}}{l_s^2k^2}  \right]~.
\end{align}
The tensor to scalar ratio is
\begin{align}
  r_{h\Phi} =  \left( 1 - \frac{T(k)}{T_\mathrm{H}}\right)^2 \ln^2 \left[ \frac{1-T(k)/T_\mathrm{H}}{l_s^2k^2}  \right]~.
\end{align}
On the other hand, the tensor to scalar ratio measured by BICEP2 is between $h$ and the comoving curvature perturbation $\zeta$. This results in a conversion factor. On super-Hubble scales, 
\begin{align}
  \zeta \simeq \left( 1 + \frac{\mathcal{H}^2}{\mathcal{H}^2-\mathcal{H}'} \right)  \Phi~,
\end{align}
where $\mathcal{H}$ is the comoving Hubble parameter and prime denotes derivative with respect to the comoving time. Assuming the Hagedorn phase is fast enough to the radiation dominated era, the factor can be calculated as
\begin{align}
  \zeta \simeq \frac{3}{2} \Phi~.
\end{align}
Thus the tensor to scalar ratio using $\zeta$ is
\begin{align}
  r = \frac{4}{9} r_{h\Phi} = 
  \frac{4}{9} \left( 1 - \frac{T(k)}{T_\mathrm{H}}\right)^2 \ln^2 \left[ \frac{1-T(k)/T_\mathrm{H}}{l_s^2k^2}  \right]~.
\end{align}
Inserting the COBE normalization $P_\zeta = 2.43 \times 10^{-9}$ and the BICEP2 central value $r=0.2$, the string scale is derived as $M_s \simeq 10^{-3} M_p$.

The tensor spectral index is
\begin{align}\label{eq:stringgas}
  n_\mathrm{T} = - (n_s -1) \left( 2 \frac{T(k)}{T_\mathrm{H}} -1 \right) \simeq -(n_s -1)~.
\end{align}
Thus string gas cosmology predicts blue tensor spectra. However, note that \eqref{eq:stringgas} is suppressed by slow roll parameters. Thus it would not explain the hint of order one $n_\mathrm{T}$ shown in data.

It is also worth to mention that string gas cosmology produces highly Gaussian density perturbations (unless the string scale is near TeV scale, which is not preferred from current data). The $f_\mathrm{NL}$ estimator for string gas cosmology can be calculated as \cite{Chen:2007js} \footnote{Note that only the non-Gaussianity near the horizon-crossing is calculated in \cite{Chen:2007js}. The Hubble-scale gravitational non-linearity may also introduce some non-Gaussianities, typically suppressed by slow varying parameters.}
\begin{align}
  f_\mathrm{NL} \sim \left(\frac{l_s}{l_p} \right) \times 10^{-30} \frac{k}{k_0}~, 
\end{align}
where $k_0$ corresponds to the scale of our present observable universe. With the detection of $r$, the string scale is known in string gas cosmology. As a result, $f_\mathrm{NL} \sim 10^{-27} k / k_0$. This is orders-of-magnitude smaller than the observational bound.

As we shall see, the inflationary candidates with blue $n_\mathrm{T}$, as far as we consider, produce considerable amount of non-Gaussianities. Thus in case a blue tensor tilt is detected, non-Gaussianity should be the next test (and maybe already much better constrained than now by the time of $n_\mathrm{T}$ measurement) to distinguish between string gas and inflationary models.

\section{Inflation: Violation of NEC}
\label{sec:infl-viol-nec}

In standard Einstein-Hilbert gravity, the amplitude of the tensor modes is determined by the energy scale of inflation. And thus for the case of blue $n_\mathrm{T}$, super inflation is needed, with null energy condition (NEC) violation \cite{Piao:2004tq}.

It has been debated for a long time that if the cosmological (inflation, alternatives or dark energy) background violates NEC, the scalar sector of the inflationary perturbations should become a ghost. However, this situation has been changed since Galileons are introduced to inflation. Here we shall explore the parameter space of the simplest G-inflation scenario, with slight generalization from the explicit model with an exponential potential in \cite{Kobayashi:2010cm} into a general slow roll functional dependence.

We start from the action of G-inflation
\begin{align}
\mathcal{L}_\phi = K(\phi, X) - G(\phi, X)\Box \phi~,
\end{align}
where $X \equiv - \frac{1}{2}\nabla_\mu \phi \nabla^\mu \phi$. Here we restrict our attention to the models with
\begin{align}
  K(\phi, X) = -X + \frac{X^2}{2M^3\mu} ~, \quad G(\phi, X) = g(\phi) X~.
\end{align}

A special class of de Sitter solution is worked out in \cite{Kobayashi:2010cm}:
\begin{align} \label{eq:gds}
  3M_p^2 H^2 = -K~, \quad K_X + 3g H \dot\phi = 0~.
\end{align}
We shall consider small derivation of the de Sitter solution \eqref{eq:gds}. For this purpose, define small parameters
\begin{align}
  \epsilon \equiv - \frac{\dot H}{H^2}~, \quad \epsilon_\phi \equiv - \frac{\ddot\phi}{H \dot\phi}~, \quad 
  \epsilon_g \equiv M_p \frac{g_\phi}{g} ~, \quad \epsilon_\mu \equiv \frac{\mu}{M_p}~.   
\end{align}
When expanding around the de Sitter solution, the $\epsilon$, $\epsilon_\phi$, and $\epsilon_g$ parameters should indeed be small. The consistency for the smallness of the $\epsilon_\mu$ parameter shall be checked later.

In terms of those slow roll parameters, and near the de Sitter solution, the order of magnitude of the following quantities can be estimated:
\begin{align}
X K_X \sim K \mathcal{O}(\epsilon_\mu)~, \quad
X g H \dot\phi \sim K \mathcal{O}(\epsilon_\mu)~, \quad
G_\phi X \sim K \mathcal{O}(\epsilon_g \epsilon_\mu)~. \quad
\end{align}

The second order action of the scalar perturbations can be written as
\begin{align}
  S^{(2)} = \frac{1}{2} \int d\tau d^3x z^2 \left[ \mathcal{G} (\mathcal{R}')^2 - \mathcal{F}(\partial_i \mathcal{R})^2 \right]~,
\end{align}
where with the above slow roll approximation, the $\mathcal{F}$ and $\mathcal{G}$ functions can be written as (for unapproximated definition, see  \cite{Kobayashi:2010cm})
\begin{align}
  \mathcal{F} = - \frac{1}{3} K_X + \mathcal{O}(\epsilon_\mu^2)~, \quad
  \mathcal{G} = 2X K_{XX} + \mathcal{O}(\epsilon_\mu)~.
\end{align}
Thus the scalar perturbations are stable in general for $g(\phi)$ satisfying the above slow roll conditions. The power spectrum and spectral index of the scalar sector can be calculated as
\begin{align} \label{eq:gscalar}
  P_\mathcal{R} = \frac{Q}{4\pi^2}~,  \quad n_s -1 = -2 \epsilon \mathcal{C}~, 
\end{align}
where
\begin{align}
  Q = \frac{K^2 }{18 M_p^4 X} \sqrt{\frac{\mathcal{G}}{\mathcal{F}^3}} ~,\quad
  \mathcal{C} = - \frac{3}{2} \frac{K K_{XX}}{K_X^2} + \mathcal{O}(\epsilon_\mu)~. 
\end{align}
Note that $K<0$ and $K_{XX}>0$ near the de Sitter solution. Thus $\mathcal{C}>0$.

The gravity sector is not modified. Thus the tensor mode has the conventional spectrum
\begin{align}
P_\mathrm{T} = \frac{2 H^2}{\pi^2 M_p^2}~.
\end{align}
The tensor spectral index is
\begin{align}
  n_\mathrm{T} = - 2\epsilon~,
\end{align}
and the tensor to scalar ratio in the slow-roll approximation is written as
\begin{equation}
  r = \frac{ 16 \sqrt{6} } {3}  \left( \sqrt{3} \epsilon_\mu \right)^{3/2}
\end{equation}
Here the tensor tilt can be blue, but the blueness is suppressed by a slow roll parameter. Also note that for the tensor mode to be blue, we need $\epsilon<0$. This is indeed possible. However, it is worth noting that from \eqref{eq:gscalar}, 
\begin{align}
  n_s - 1 > 0~.
\end{align}
In other words, near the de Sitter solution \eqref{eq:gds}, the scalar and tensor modes tilt towards the same direction -- both red spectra or both blue spectra. Note that a blue scalar spectrum is not favored by observations. However, in more general cosmological models, for example, with a different number of neutrinos, a blue scalar spectrum is not yet ruled out.

On the other hand, we are not sure at this point if the same direction of tilt for scalar and tensor spectra is a general feature of G-inflation (and generalized Galileons \cite{Deffayet:2009mn, Deffayet:2011gz, Kobayashi:2011nu}), or there exists unexplored models with different tilt of two sectors.

Also, one has to note the sound speed of the scalar sector of perturbations is
\begin{align}
  c_s^2 = \frac{\mathcal{F}}{\mathcal{G}} = - \frac{1}{6} \frac{K_X}{X K_{XX}} \sim \frac{\epsilon_\mu}{ 2 \sqrt{3}}
\end{align}
This would induce an equilateral non-Gaussianity $|f^\mathrm{equil}_\mathrm{NL}|\sim 1/c_s^2$. This is consistent with the current \textit{Planck} bound, and on the other hand accessible in the future. 

It is worth mentioning that the small $c_s$ and large $|f^\mathrm{equil}_\mathrm{NL}|$ are not a coincidence in this particular toy model, but should be a rather model-independent statement (unless fine tuned). To violate NEC at the background level while keeping the perturbations stable, the relevant Galileons Lagrangian at the energy scale of inflation should be highly non-linear. Non-linear self-coupling introduces equilateral non-Gaussianity. Experiments in the near future would reduce the bound for $|f^\mathrm{equil}_\mathrm{NL}|$ (or detection) and examine the possibility of Galileons and NEC violation.

Finally, one can check that when $r\simeq 0.2$, $\epsilon_\mu \simeq 0.03$. Thus the smallness of $\epsilon_\mu$ is consistent with data.

\section{Inflation: General initial conditions}
\label{sec:infl-gener-init}

It has been an open question if the inflationary (scalar and tensor) perturbations originates from a Bunch-Davies (BD) vacuum, or special care needs to be taken for the non-BD initial conditions. One motivation for choosing non-BD initial conditions, for example, is the transPlanckian problem of inflation \cite{Brandenberger:2000wr, Martin:2000xs} (see \cite{Brandenberger:2012aj} for a recent review) \footnote{There are also other possibilities. For example, thermal fluctuations as initial condition of inflation \cite{Biswas:2013lna}.}. 

\begin{figure}[htbp]
  \centering
  \includegraphics[width=0.8\textwidth]{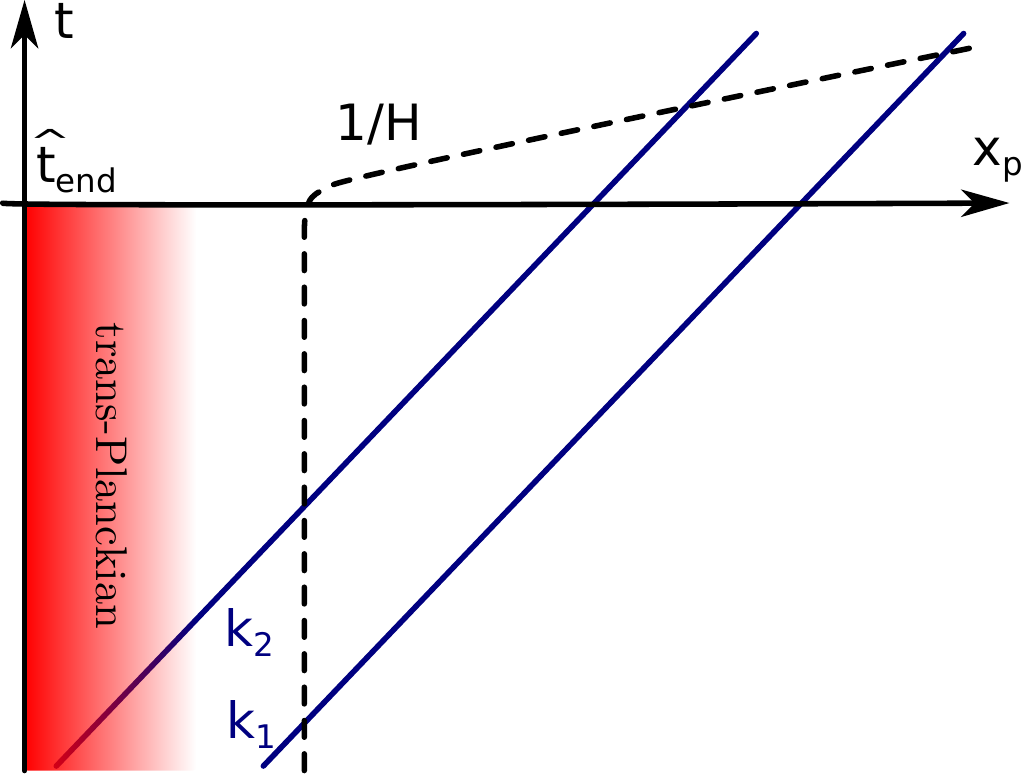}
  \caption{\label{fig:trans} The transPlanckian problem. The inflationary perturbations are initialized on sub-Planckian scales (or above other UV completion scales).}
\end{figure}

Inflation requires UV completion. During inflation, the perturbations originates from the UV completion scale, for example, Planck scale, before they expand and cross the Hubble horizon. Note that the Planck scale could have been replaced by some lower scales, for example, the string scale. With those energy scales for new physics, we can no longer make sure that the inflationary perturbations were in their lowest energy states before they got stretched by the cosmic expansion. This situation is illustrated in Fig.~\ref{fig:trans}.

In the context of tensor perturbations \cite{Ashoorioon:2013eia} (for a related recent study, see also \cite{Collins:2014yua}), the action of tensor mode is 
\begin{align}
  S = \frac{M_p^2}{8} \int \frac{d^3k}{(2\pi)^3}d\tau~a^2
  \left( \gamma^i_j{}' \gamma^j_i{}' - k^2 \gamma^i_j \gamma^j_i\right) ~.
\end{align}
Decomposing the tensor modes by polarization, we have
\begin{align} \label{eq:tensor-dec}
  \gamma_{ij}(\mathbf{k}) = \frac{\sqrt2}{M_p} \left[ \gamma_+(\mathbf{k}) e^+_{ij}(\mathbf{k}) 
    + \gamma_\times (\mathbf{k}) e^\times_{ij}(\mathbf{k}) \right]~,
\end{align}
as a result, the action contains two copies of modes
\begin{align}
  S = \frac{1}{2} \int \frac{d^3k}{(2\pi)^3}d\tau~a^2
  \left[ \left( \gamma_+' \gamma_+{}' - k^2 \gamma_+ \gamma_+\right) 
    + \left( \gamma_\times' \gamma_\times{}' - k^2 \gamma_\times \gamma_\times\right) \right]~.
\end{align}
For each mode, one can impose a different initial condition from BD. 

The $\gamma$ (collectively denote $\gamma_+$ and $\gamma_\times$) field can be quantized as
\begin{align}
  \gamma_\mathrm{k} = v_k a_\mathrm{k} + v^*_k a^\dagger_{-\mathrm{k}}
\end{align}
where 
\begin{align}
v_k = C_+(k) \frac{H}{\sqrt{2k^3}} (1+i k \tau)e^{-ik\tau} + C_-(k) \frac{H}{\sqrt{2k^3}} (1-i k \tau)e^{ik\tau}~. 
\end{align}
The quantization condition requires
\begin{align}
  |C_+(k)|^2 - |C_-(k)|^2 = 1~. 
\end{align}
The lowest energy state has $C_=0$, which corresponds to the BD state.

The tensor power spectrum for each polarization mode is
\begin{align}
P^{+,\times}_\gamma = |C_+^{+,\times}(k)+C_-^{+,\times}(k)|^2 \left( \frac{H}{2\pi}  \right)^2~.  
\end{align}

\begin{align}
P_\gamma = \frac{2H^2}{\pi^2M_p^2} \left[ |C_+^{+}(k)+C_-^{+}(k)|^2+|C_+^{\times}(k)+C_-^{\times}(k)|^2 \right] ~.  
\end{align}
Note that the $k$-dependence of $C_{+,-}^{+,\times}(k)$ is determined by high energy physics and is unknown at low energy scales. Thus this $k$-dependence could in principle tilt the tensor spectrum to the blue end and overwrite the red tilt from decreasing $H$. This requires
\begin{align}
  \frac{\partial}{\partial\ln k} \left\{ H^2  \left[ |C_+^{+}(k)+C_-^{+}(k)|^2+|C_+^{\times}(k)+C_-^{\times}(k)|^2 \right]\right\} >0~.
\end{align}
and the tensor tilt is
\begin{align}
  n_\mathrm{T} = -2 \epsilon + 
  \frac{d\ln \left[ |C_+^{+}(k)+C_-^{+}(k)|^2+|C_+^{\times}(k)+C_-^{\times}(k)|^2 \right]}{d \ln k}~. 
\end{align}

We would like to make a few comments before closing up this section:
\begin{itemize}
\item The non-BD state of tensor modes also opens up possibilities for small field inflation to be consistent with the observed large tensor to scalar ratio $r=0.2$. However, note that a reason is needed for the scalar sector and the tensor sector to achieve different non-BD coefficients. 
\item Non-BD initial condition generically introduce non-Gaussianities of the folded shape \cite{Chen:2006nt,Xue:2008mk}. The size of the non-Gaussianity is proportional to $\mathrm{Re}[C^-(k)]$ and peaked at the folded limit $\mathbf{k}_1+\mathbf{k}_2=\mathbf{k}_3$. In the squeezed limit $k_3\rightarrow 0$, the folded non-Gaussianity also blows up more quickly than local shape (however, in the regime under theoretical control, the squeezed limit remains small \cite{Flauger:2013hra,Aravind:2013lra}). Thus the tensor modes generated by non-BD coefficients are non-Gaussian. Despite of the non-Gaussianity in the tensor sector, it is also reasonable to expect a folded non-Gaussianity for the scalar sector. Because if the non-BD issue exists, it is likely not tensor-only unless a reason is provided.
\end{itemize}

\section{Other possibilities}
\label{sec:other-possibilities}

There is a huge landscape of inflation models, and also quite a few alternatives to inflation. Some of them may also be able to produce blue $n_\mathrm{T}$. We list a few possibilities here.

\subsection{Inflation: External sources for tensor modes}
Besides gravitational waves from quantum fluctuations during inflation, 
there is a possibility that the dominant gravitational waves are 
generated by particles or strings \cite{Senatore:2011sp}, or particle states \cite{Cook:2011hg, Carney:2012pk}  produced in the period inflation. The kinetic
energy of the inflaton $\dot{\phi}^2 \sim 2\epsilon H^2 M_p^2$ dumps
to the particles, and it is large enough to produce visible tensor spectrum.
The emission rate is related to the square of the coupling constant,
i.e, for the particles with energy $E$, $\mathrm{rate} \propto E^2 / M_p^2$.

One can derive a bound on $\epsilon$ from those class of particle production mechanism of sourcing tensor mode. One can estimate the energy density of the classically sourced gravitational waves energy density $\rho_\mathrm{GW}$ near horizon crossing:
\begin{align}
  \rho_\mathrm{GW} < \epsilon M_p^2 H^2~,
\end{align} 
where $\epsilon M_p^2 H^2$ is the kinetic energy density of the inflaton. On the other hand, near horizon crossing, the physical wave length of the tensor modes is of order Hubble scale. Thus
\begin{align}
  \rho_\mathrm{GW} \sim M_p^2 (\partial h)^2 \sim M_p^2 H^2 h^2~.
\end{align}
Thus the tensor power spectrum
\begin{align}
  P_\mathrm{T} \sim \langle h^2\rangle < \epsilon~.
\end{align}
From the observed tensor to scalar ratio
\begin{align}
  0.2 = r < \frac{\epsilon}{P_\zeta} = \frac{\epsilon}{2\times 10^{-9}}~,  
\end{align}
We get
\begin{align}
  \epsilon > 4 \times 10 ^{-10}~.
\end{align}
From COBE normalization, we get
\begin{align}
  H > 9\times 10^{-9}M_p \sim 2 \times 10 ^{10} \mathrm{GeV}~, \qquad \rho = 3M_p^2H^2 > 2 \times 10^{-16}M_p^4 \sim (3 \times 10^{14}  \mathrm{GeV})^4
\end{align}
Below those scales, no classical source could generate primordial tensor modes.

In some simple models proposed in \cite{Senatore:2011sp}, the produced particles have a time-dependent mass.
$M^2 \sim \dot\phi^2 t^2$ due to its coupling to the rolling of inflaton.
Compared with the fact that inflation has the decreasing Hubble constant, which
predict a red tensor spectrum,
the increasing mass will afford an explanation of the blue tensor spectrum.

It is also important to note that non-Gaussianities are also produced from the particle production process \cite{Barnaby:2012xt}. The constraints from non-Gaussianity of particle production remains to be tested.

\subsection{Inflation: Beyond slow roll}

Based on the slow roll expansion \cite{Gong:2004kd}, it is recently pointed out that violation of slow roll can generate blue tensor spectrum \cite{Gong:2014qga}. In \cite{Gong:2004kd}, the tensor spectral index is calculated up to second order in slow roll as
\begin{align}
  n_\mathrm{T} \simeq -2 \epsilon - 2 \epsilon^2 - 0.54 \epsilon\eta~,
\end{align}
Thus if the slow roll expansion is extrapolated to $\eta \lesssim -3.7$, blue tensor spectrum is obtained. For example, the possible blue tensor spectrum in the ultra-slow-roll inflation \cite{Tsamis:2003px, Chen:2013aj} is discussed in \cite{Gong:2014qga}. It would be interesting to further examine this possibility without slow roll restrictions, and see if the introduced large running is consistent with data. 



\subsection{Inflation: Modified gravity}

The tensor modes come from the gravity sector. Thus there is no surprise that if gravity is modified, the tensor spectrum could change.

Among other possibilities, massive gravity is one example to achieve this goal. Massive gravity on a time dependent background is a challenge \cite{D'Amico:2011jj}. Nevertheless, there exist viable models \cite{Gumrukcuoglu:2011ew, DeFelice:2013tsa, DeFelice:2013dua} and application to inflation \cite{Dubovsky:2009xk, Lin:2013sja}. We may expect tensor modes with blue spectra in massive gravity: If $m^2>0$ for the graviton during inflation, the tensor modes which exit the horizon earlier have more time to roll back to the origin of their mass potential, and thus are suppressed more. As a result, a blue tensor tilt can be generated.

However, the massive gravity in the late universe does not help for the tensor modes. This is because the mass of the graviton is too small. Also, the return-to-horizon effect actually generates a redder spectrum for the gravitational waves \cite{Gumrukcuoglu:2012wt}. 

\subsection{Inflation: With space-like condensates}

As another interesting class of possibilities, some inflation models come with space-like condensates of fields. For example, space-like gradient of scalars in solid inflation \cite{Endlich:2012pz} and SO(3) massive gravity \cite{Lin:2013sja}, or space-like vector fields in Chromo-Natural inflation \cite{Adshead:2012kp}. Typically, SO(3) spatial rotational symmetry is imposed to keep isotropy. In this class of models, there can be additional tensor components from decomposition of the SO(3) group, other than the graviton. Those components can also contribute to the B-mode power spectrum. For the additional components, there is no NEC type restriction against blue spectra. For example, solid inflation predicts a slightly blue tensor spectrum \cite{Endlich:2012pz, Akhshik:2014gja}. Note that the blue tilt in solid inflation is on the one hand suppressed by $\epsilon$ but on the other hand boosted by sound speed. For $r=0.2$ one finds $n_\mathrm{T} = 0.065$.

\subsection{Second order effects}

With the assumption of isotropy \footnote{Beyond such an assumption, in anisotropic inflation it is possible to source tensor modes linearly from scalar modes \cite{Chen:2014eua}.}, scalar modes cannot source the tensor modes at the linear level. However, beyond the linear level, tensor modes can be sourced by scalar modes \cite{Matarrese:1997ay}. Those second order effects are typically unobservably small because they are suppressed by the inflationary power spectrum. However, the second order contribution to tensor mode can be boosted by a small sound speed of an isocurvature scalar sector \cite{Biagetti:2013kwa}. The generated gravitational waves are highly non-Gaussian. The tilt of tensor spectrum coming from those contributions are not constrained by the null energy condition, thus a blue tilt is possible.

\subsection{Matter bounce}

As another alternative to inflation scenario, in matter bounce, the universe was in a contracting phase before bouncing back and heating up (see \cite{Brandenberger:2009jq} for a review). In matter bounce, the tensor modes are nearly scale invariant but the amplitudes have been too high. Recently, a two-field bounce model is introduced and the tensor modes can be tuned as a parameter \cite{Cai:2013kja}. It is thus interesting to see if this model (among other possibilities) fit the current observations and has a potential for blue tensor spectrum.

\section{Conclusion and discussion}
\label{sec:concl-disc}

To conclude, we fit the tensor-to-scalar ratio and the tensor spectral tilt with data. The current data is not good enough to test the inflationary consistency relation but nevertheless blue tensor spectra are favored when all 9 bins of BICEP2 data, in combination with the POLARBEAR data, are used. 

From theoretical aspects, string gas cosmology predicts blue tensor spectra. However, the tilt is small, at the same order-of-magnitude of scalar tilt. Thus the future experiments targeting to test the inflationary consistency relation can also test this prediction of string gas cosmology. On the other hand, string gas cosmology predicts highly Gaussian density and tensor perturbations. This is unlike the other inflationary mechanisms, where a blue tilt also implies non-Gaussianities.

The simplest model of G-inflation, on the other hand, tilts scalar and tensor power spectra in the same way. Thus with a red scalar spectral tilt, the tensor spectra are also red. In the parameter regime where both scale and tensor spectra are blue, the blueness of tensor perturbation is suppressed by slow roll parameter $\epsilon$. The equilateral non-Gaussianity of G-inflation, at $r=0.2$, is about $30$ and close to the current observational bound. The non-Gaussianities should be a model independent feature for super-inflation type models which generate blue tensor spectrum, because the perturbations have to behave differently from the background to avoid ghosts.

Generalized initial conditions of inflation is left largely unconstrained. However, the generalized initial conditions are also sources of non-Gaussianities. Thus non-Gaussianities in the tensor sector would be a test of those class of models.

Inflationary particle production is another possible source of tensor modes. We derived a bound of inflationary Hubble scale and energy density for this mechanism to work. Non-Gaussianities are also present in the case of particle productions. We hope to investigate the particle production mechanism and its relation between the blue tensor spectra in a future work. 

The current data shows hint of blue $n_\mathrm{T} \sim \mathcal{O}(1)$. The models with slow roll suppressed $n_\mathrm{T}$ would not be enough to explain such a hint. On the other hand, from modified initial conditions, external sources, inflation beyond slow roll, modified gravity and some models of matter bounce, blue and large $n_\mathrm{T}$ may be produced. It would be interesting to study those models in more details to see how the fitting of data is improved in those models.

\noindent\textbf{Note added:}

Two related works \cite{Gerbino:2014eqa, Ashoorioon:2014nta} appeared on arXiv on the same day as ours. \cite{Gerbino:2014eqa} (see also \cite{cosmocoffee}) overlaps with our Section \ref{sec:hints-blue-tensor}, and \cite{Ashoorioon:2014nta} overlaps with Section \ref{sec:infl-gener-init}. 

The data analysis of tensor tilt is also investigated by \cite{Cheng:2014bma}, \cite{Wu:2014qxa} and \cite{Cheng:2014ota}, within a few days before/after our paper. Among those papers, \cite{Cheng:2014bma} and \cite{Cheng:2014ota} reports a nearly zero central value, with small errorbars $\Delta n_\mathrm{T} \sim 0.48$ (BICEP2 only) and $\Delta n_\mathrm{T} \sim 0.24$ (BICEP2 + Planck + WP, with running of scalar spectral index). While the following works prefer blue tilt with considerably larger errorbars: \cite{Gerbino:2014eqa} (BICEP2 only), our result (BICEP2, BICEP2 + Planck + WP, POLARBEAR), \cite{cosmocoffee} (BICEP2 + Planck + WP and BICEP2 x Keck + Planck + WP) and \cite{Wu:2014qxa} (BICEP2 + Planck + WMAP + BAO).

\section*{Acknowledgments}

YW is supported by a Starting Grant of the European Research Council (ERC STG grant 279617), and the Stephen Hawking Advanced Fellowship. This work was undertaken on the COSMOS Shared Memory system at DAMTP, University of Cambridge operated on behalf of the STFC DiRAC HPC Facility. This equipment is funded by BIS National E-infrastructure capital grant ST/J005673/1 and STFC grants ST/H008586/1, ST/K00333X/1.



\begin{thebibliography}{999}

\bibitem{Guth81} A.~H.~Guth, Phys. Rev. D. 23 (1981) 347

\bibitem{Linde82} A.~D.~Linde, Phys. Lett. B. 108 (1982) 38

\bibitem{Planck16} P.~A.~P.~Ade et al., Planck 2013 results XVI., arXiv: 1303.5076

\bibitem{Ade:2014xna} 
  P.~A.~R.~Ade {\it et al.}  [BICEP2 Collaboration],
  arXiv:1403.3985 [astro-ph.CO].

\bibitem{Mortonson:2014bja} 
  M.~J.~Mortonson and U.~Seljak,
  arXiv:1405.5857 [astro-ph.CO].

\bibitem{Flauger:2014qra} 
  R.~Flauger, J.~C.~Hill and D.~N.~Spergel,
  JCAP {\bf 1408}, 039 (2014)
  [arXiv:1405.7351 [astro-ph.CO]].

\bibitem{Copeland:1994vg} 
  E.~J.~Copeland, A.~R.~Liddle, D.~H.~Lyth, E.~D.~Stewart and D.~Wands,
  Phys.\ Rev.\ D {\bf 49}, 6410 (1994)
  [astro-ph/9401011].

\bibitem{Ma:2014vua} 
  Y.~-Z.~Ma and Y.~Wang,
  arXiv:1403.4585 [astro-ph.CO].

\bibitem{Planck22} P.~A.~P.~Ade et al., Planck 2013 results XXII,
arXiv: 1303.5082

\bibitem{Ade:2014afa} 
  P.~A.~R.~Ade {\it et al.}  [ The POLARBEAR Collaboration],
  arXiv:1403.2369 [astro-ph.CO].

\bibitem{Austermann12} J.~E.~Austermann et al., 2012, SPIE, 8452, 1

\bibitem{Niemack10} N.~D.~Niemack et al., 2010, SPIE, 7741, 51

\bibitem{Eimer12} J.~R.~Eimer et al., 2012, SPIE, 8452, 20

\bibitem{Lewis:2002ah} 
  A.~Lewis and S.~Bridle,
  Phys.\ Rev.\ D {\bf 66}, 103511 (2002)
  [astro-ph/0205436].

\bibitem{Blas:2011rf}
  D.~Blas, J.~Lesgourgues and T.~Tram,
  JCAP {\bf 1107} (2011) 034
  [arXiv:1104.2933 [astro-ph.CO]].

\bibitem{Hinshaw:2012aka}
  G.~Hinshaw {\it et al.}  [WMAP Collaboration],
  Astrophys.\ J.\ Suppl.\  {\bf 208} (2013) 19
  [arXiv:1212.5226 [astro-ph.CO]].

\bibitem{Ashoorioon:2014nta} 
  A.~Ashoorioon, K.~Dimopoulos, M.~M.~Sheikh-Jabbari and G.~Shiu,
  arXiv:1403.6099 [hep-th].

\bibitem{Ade:2013kta} 
  P.~A.~R.~Ade {\it et al.}  [Planck Collaboration],
  arXiv:1303.5075 [astro-ph.CO].

\bibitem{Hanson:2013hsb} 
  D.~Hanson {\it et al.}  [SPTpol Collaboration],
  Phys.\ Rev.\ Lett.\  {\bf 111}, 141301 (2013)
  [arXiv:1307.5830 [astro-ph.CO]].

\bibitem{Cai:2014hja} 
  Y.~-F.~Cai and Y.~Wang,
  arXiv:1404.6672 [astro-ph.CO].

\bibitem{Brandenberger:1988aj} 
  R.~H.~Brandenberger and C.~Vafa,
  Nucl.\ Phys.\ B {\bf 316}, 391 (1989).

\bibitem{Nayeri:2005ck} 
  A.~Nayeri, R.~H.~Brandenberger and C.~Vafa,
  Phys.\ Rev.\ Lett.\  {\bf 97}, 021302 (2006)
  [hep-th/0511140].

\bibitem{Brandenberger:2006xi} 
  R.~H.~Brandenberger, A.~Nayeri, S.~P.~Patil and C.~Vafa,
  Phys.\ Rev.\ Lett.\  {\bf 98}, 231302 (2007)
  [hep-th/0604126].

\bibitem{Brandenberger:2006vv} 
  R.~H.~Brandenberger, A.~Nayeri, S.~P.~Patil and C.~Vafa,
  Int.\ J.\ Mod.\ Phys.\ A {\bf 22}, 3621 (2007)
  [hep-th/0608121].

\bibitem{Brandenberger:2014faa} 
  R.~H.~Brandenberger, A.~Nayeri and S.~P.~Patil,
  arXiv:1403.4927 [astro-ph.CO].

\bibitem{Khoury:2001wf} 
  J.~Khoury, B.~A.~Ovrut, P.~J.~Steinhardt and N.~Turok,
  Phys.\ Rev.\ D {\bf 64}, 123522 (2001)
  [hep-th/0103239].

\bibitem{Biswas:2014kva} 
  T.~Biswas, T.~Koivisto and A.~Mazumdar,
  JHEP {\bf 1408}, 116 (2014)
  [arXiv:1403.7163 [hep-th]].

\bibitem{Boyle:2003km} 
  L.~A.~Boyle, P.~J.~Steinhardt and N.~Turok,
  Phys.\ Rev.\ D {\bf 69}, 127302 (2004)
  [hep-th/0307170].

\bibitem{Qiu:2011cy} 
  T.~Qiu, J.~Evslin, Y.~-F.~Cai, M.~Li and X.~Zhang,
  JCAP {\bf 1110}, 036 (2011)
  [arXiv:1108.0593 [hep-th]].

\bibitem{Chen:2007js} 
  B.~Chen, Y.~Wang, W.~Xue and R.~Brandenberger,
  arXiv:0712.2477 [hep-th].

\bibitem{Piao:2004tq} 
  Y.~-S.~Piao and Y.~-Z.~Zhang,
  Phys.\ Rev.\ D {\bf 70}, 063513 (2004)
  [astro-ph/0401231].

\bibitem{Kobayashi:2010cm} 
  T.~Kobayashi, M.~Yamaguchi and J.~'i.~Yokoyama,
  Phys.\ Rev.\ Lett.\  {\bf 105}, 231302 (2010)
  [arXiv:1008.0603 [hep-th]].

\bibitem{Deffayet:2009mn} 
  C.~Deffayet, S.~Deser and G.~Esposito-Farese,
  Phys.\ Rev.\ D {\bf 80}, 064015 (2009)
  [arXiv:0906.1967 [gr-qc]].

\bibitem{Deffayet:2011gz} 
  C.~Deffayet, X.~Gao, D.~A.~Steer and G.~Zahariade,
  Phys.\ Rev.\ D {\bf 84}, 064039 (2011)
  [arXiv:1103.3260 [hep-th]].

\bibitem{Kobayashi:2011nu} 
  T.~Kobayashi, M.~Yamaguchi and J.~'i.~Yokoyama,
  Prog.\ Theor.\ Phys.\  {\bf 126}, 511 (2011)
  [arXiv:1105.5723 [hep-th]].

\bibitem{Brandenberger:2000wr} 
  R.~H.~Brandenberger and J.~Martin,
  Mod.\ Phys.\ Lett.\ A {\bf 16}, 999 (2001)
  [astro-ph/0005432].

\bibitem{Martin:2000xs} 
  J.~Martin and R.~H.~Brandenberger,
  Phys.\ Rev.\ D {\bf 63}, 123501 (2001)
  [hep-th/0005209].

\bibitem{Brandenberger:2012aj} 
  R.~H.~Brandenberger and J.~Martin,
  Class.\ Quant.\ Grav.\  {\bf 30}, 113001 (2013)
  [arXiv:1211.6753 [astro-ph.CO]].

\bibitem{Biswas:2013lna} 
  T.~Biswas, R.~Brandenberger, T.~Koivisto and A.~Mazumdar,
  Phys.\ Rev.\ D {\bf 88}, no. 2, 023517 (2013)
  [arXiv:1302.6463 [astro-ph.CO]].

\bibitem{Ashoorioon:2013eia} 
  A.~Ashoorioon, K.~Dimopoulos, M.~M.~Sheikh-Jabbari and G.~Shiu,
  JCAP {\bf 1402}, 025 (2014)
  [arXiv:1306.4914 [hep-th]].

\bibitem{Collins:2014yua} 
  H.~Collins, R.~Holman and T.~Vardanyan,
  arXiv:1403.4592 [hep-th].

\bibitem{Chen:2006nt} 
  X.~Chen, M.~-x.~Huang, S.~Kachru and G.~Shiu,
  JCAP {\bf 0701}, 002 (2007)
  [hep-th/0605045].

\bibitem{Xue:2008mk}
  W.~Xue and B.~Chen,
  Phys.\ Rev.\ D {\bf 79} (2009) 043518
  [arXiv:0806.4109 [hep-th]].

\bibitem{Flauger:2013hra} 
  R.~Flauger, D.~Green and R.~A.~Porto,
  JCAP {\bf 1308}, 032 (2013)
  [arXiv:1303.1430 [hep-th]].

\bibitem{Aravind:2013lra} 
  A.~Aravind, D.~Lorshbough and S.~Paban,
  JHEP {\bf 1307}, 076 (2013)
  [arXiv:1303.1440 [hep-th]].

\bibitem{Senatore:2011sp}
  L.~Senatore, E.~Silverstein and M.~Zaldarriaga,
  arXiv:1109.0542 [hep-th].

\bibitem{Cook:2011hg} 
  J.~L.~Cook and L.~Sorbo,
  Phys.\ Rev.\ D {\bf 85}, 023534 (2012)
  [Erratum-ibid.\ D {\bf 86}, 069901 (2012)]
  [arXiv:1109.0022 [astro-ph.CO]].

\bibitem{Carney:2012pk} 
  D.~Carney, W.~Fischler, E.~D.~Kovetz, D.~Lorshbough and S.~Paban,
  JHEP {\bf 1211}, 042 (2012)
  [arXiv:1209.3848 [hep-th]].

\bibitem{Barnaby:2012xt} 
  N.~Barnaby, J.~Moxon, R.~Namba, M.~Peloso, G.~Shiu and P.~Zhou,
  Phys.\ Rev.\ D {\bf 86}, 103508 (2012)
  [arXiv:1206.6117 [astro-ph.CO]].

\bibitem{Gong:2004kd} 
  J.~-O.~Gong,
  Class.\ Quant.\ Grav.\  {\bf 21}, 5555 (2004)
  [gr-qc/0408039].

\bibitem{Gong:2014qga} 
  J.~-O.~Gong,
  arXiv:1403.5163 [astro-ph.CO].

\bibitem{Tsamis:2003px} 
  N.~C.~Tsamis and R.~P.~Woodard,
  Phys.\ Rev.\ D {\bf 69}, 084005 (2004)
  [astro-ph/0307463].

\bibitem{Chen:2013aj} 
  X.~Chen, H.~Firouzjahi, M.~H.~Namjoo and M.~Sasaki,
  Europhys.\ Lett.\  {\bf 102}, 59001 (2013)
  [arXiv:1301.5699 [hep-th]].

\bibitem{D'Amico:2011jj} 
  G.~D'Amico, C.~de Rham, S.~Dubovsky, G.~Gabadadze, D.~Pirtskhalava and A.~J.~Tolley,
  Phys.\ Rev.\ D {\bf 84}, 124046 (2011)
  [arXiv:1108.5231 [hep-th]].

\bibitem{Gumrukcuoglu:2011ew} 
  A.~E.~Gumrukcuoglu, C.~Lin and S.~Mukohyama,
  JCAP {\bf 1111}, 030 (2011)
  [arXiv:1109.3845 [hep-th]].

\bibitem{DeFelice:2013tsa} 
  A.~De Felice and S.~Mukohyama,
  Phys.\ Lett.\ B {\bf 728}, 622 (2014)
  [arXiv:1306.5502 [hep-th]].

\bibitem{DeFelice:2013dua} 
  A.~De Felice, A.~E.~Gumrukcuoglu and S.~Mukohyama,
  Phys.\ Rev.\ D {\bf 88}, 124006 (2013)
  [arXiv:1309.3162 [hep-th]].

\bibitem{Dubovsky:2009xk} 
  S.~Dubovsky, R.~Flauger, A.~Starobinsky and I.~Tkachev,
  Phys.\ Rev.\ D {\bf 81}, 023523 (2010)
  [arXiv:0907.1658 [astro-ph.CO]].

\bibitem{Lin:2013sja} 
  C.~Lin,
  arXiv:1307.2574.

\bibitem{Gumrukcuoglu:2012wt} 
  A.~E.~Gumrukcuoglu, S.~Kuroyanagi, C.~Lin, S.~Mukohyama and N.~Tanahashi,
  Class.\ Quant.\ Grav.\  {\bf 29}, 235026 (2012)
  [arXiv:1208.5975 [hep-th]].

\bibitem{Endlich:2012pz} 
  S.~Endlich, A.~Nicolis and J.~Wang,
  JCAP {\bf 1310}, 011 (2013)
  [arXiv:1210.0569 [hep-th]].

\bibitem{Akhshik:2014gja} 
  M.~Akhshik, R.~Emami, H.~Firouzjahi and Y.~Wang,
  arXiv:1405.4179 [astro-ph.CO].

\bibitem{Adshead:2012kp} 
  P.~Adshead and M.~Wyman,
  Phys.\ Rev.\ Lett.\  {\bf 108}, 261302 (2012)
  [arXiv:1202.2366 [hep-th]].

\bibitem{Chen:2014eua} 
  X.~Chen, R.~Emami, H.~Firouzjahi and Y.~Wang,
  JCAP {\bf 1408}, 027 (2014)
  [arXiv:1404.4083 [astro-ph.CO]].

\bibitem{Matarrese:1997ay} 
  S.~Matarrese, S.~Mollerach and M.~Bruni,
  Phys.\ Rev.\ D {\bf 58}, 043504 (1998)
  [astro-ph/9707278].

\bibitem{Biagetti:2013kwa} 
  M.~Biagetti, M.~Fasiello and A.~Riotto,
  Phys.\ Rev.\ D {\bf 88}, no. 10, 103518 (2013)
  [arXiv:1305.7241 [astro-ph.CO]].

\bibitem{Brandenberger:2009jq} 
  R.~H.~Brandenberger,
  Int.\ J.\ Mod.\ Phys.\ Conf.\ Ser.\  {\bf 01}, 67 (2011)
  [arXiv:0902.4731 [hep-th]].

\bibitem{Cai:2013kja} 
  Y.~-F.~Cai, E.~McDonough, F.~Duplessis and R.~H.~Brandenberger,
  JCAP {\bf 1310}, 024 (2013)
  [arXiv:1305.5259 [hep-th]].

\bibitem{Gerbino:2014eqa} 
  M.~Gerbino, A.~Marchini, L.~Pagano, L.~Salvati, E.~Di Valentino and A.~Melchiorri,
  arXiv:1403.5732 [astro-ph.CO].

\bibitem{cosmocoffee}
Antony Lewis, http://cosmocoffee.info/viewtopic.php?t=2302.

\bibitem{Cheng:2014bma} 
  C.~Cheng and Q.~-G.~Huang,
  arXiv:1403.5463 [astro-ph.CO].

\bibitem{Wu:2014qxa} 
  F.~Wu, Y.~Li, Y.~Lu and X.~Chen,
  arXiv:1403.6462 [astro-ph.CO].

\bibitem{Cheng:2014ota} 
  C.~Cheng and Q.~-G.~Huang,
  arXiv:1403.7173 [astro-ph.CO].

\end{thebibliography}
\end{document}